\documentclass[a4paper,11pt,oneside]{article}
\usepackage[english]{babel}
\usepackage[dvips]{graphicx}
\usepackage[svgnames]{xcolor}
 
\usepackage{amssymb}
\usepackage[tbtags]{amsmath}
\usepackage{amsthm}

\usepackage[T1]{fontenc}
\usepackage{lmodern}

\usepackage[top=32mm]{geometry}

\usepackage{caption}
\usepackage{fancyhdr}
\usepackage{hyperref}
\hypersetup{
  linktocpage=true,
  linkcolor  = MidnightBlue,
  citecolor  = MidnightBlue,
  urlcolor   = MidnightBlue,
  colorlinks = true,
}

\title{Identification of independent patterns of COVID-19 mortality in Brazil by a functional QR decomposition\thanks{Preprint of an article submitted for consideration in Biophysical Reviews and Letters. \copyright~2022 World Scientific Publishing Company, https://www.worldscientific.com/worldscinet/brl}}
\author{Jorge C. Lucero\thanks{Dept.\ Computer Science, University of Bras\'{i}lia, Brazil. E-mail: \href{mailto:lucero@unb.br}{lucero@unb.br} }} 
\date{\today}

\begin{document}

\maketitle

\begin{abstract}
The subset selection problem of linear algebra is applied to identify independent patterns of COVID-19 evolution within Brazil. The data consist of a set of mortality curves in states of Brazil. A subset of the most independent curves is selected by using a functional version of the QR matrix decomposition technique with column pivoting. The selected subset is used next as a basis to represent the remaining curves filtering out any data redundancy. For each independent curve, an associated epidemiological region of influence is defined. The results show two main independent curves with a similar two-peak pattern and a 50-day shift between the patterns. Two main epidemiological regions are next identified: one encompassing most of the country from the center and northeast states to the south, an another one containing the Amazonian region at the northwest.
\end{abstract}

\section{Introduction}

\label{intro}

On March 11, 2020, the World Health Organization declared a worldwide pandemic of the COVID-19 disease caused by the new coronavirus SARS-CoV-2. The disease was identified for the first time in Wuhan, People's Republic of China, in December 2019. As of the present date (January 21, 2022), around 344 million cases have been reported, with 5.6 million deaths \cite{Worldometer2020}. In Brazil, the number of cases reaches 23 millions, and over 622 thousands lives have been lost \cite{Health2020}.

Throughout the world, health authorities have implemented vaccination campaigns and a number of measures enforcing adequate hygiene and social distancing \cite{Mathieu2021,10.1093/jtm/taaa020}. Naturally, the response to those campaigns and measures depends on demographic characteristics, compliance of the population, timing, and emergence of mutations of the virus \cite{di2021evaluation,sabino2021resurgence}. Thus, data-driven models of the pandemic propagation constitute a useful tool to characterize and analyze underlying patterns, assess the effectiveness of implemented policies, forecast its evolution, and a number of them have been proposed \cite{kuhl2020data}.

Here, we consider a modeling approach based on the QR decomposition technique of linear algebra \cite{Golub1996}, in order to identify regions with independent patterns of COVID-19 evolution within Brazil. The QR decomposition is a matrix factorization technique that provides a simple and numerically robust solution to the so-called ``subset selection problem''. In that problem, a set of observations $n$ vectors is given and a subset of the $k$ most independent ones is sought. The subset may be used next as a basis to represent the $n-k$ remaining vectors filtering out any data redundancy. This process has some similarities to the well-known technique of principal component analysis (PCA), in the sense that it achieves a reduction of the dimensionality of the data. However, instead of expressing the data in terms of transformations of the data, it does so in terms of a set of the most nonredundant observation vectors and therefore the results tend to have an easier interpretation \cite{cuevas2014partial}.

In a previous study \cite{Lucero2008a}, the QR decomposition was applied to identify kinematic regions of the face that follow independent motion patterns during speech. The study argued that, whereas PCA could be used to extract facial gestures (i.e., temporal patterns of motion), the QR decomposition approach was more adequate to express the motion of the face in terms of eigenregions which acted as independent biomechanical units. The present study has a similar purpose in the sense that it intends to build a spatio-temporal model in terms of regions of independent behavior. Therefore, the same modeling strategy of the previous facial study  will be followed, except that a functional extension of the QR decomposition will be considered. 

The proposed extension fits within a functional data analysis (FDA) context \cite{Ramsay1997}, in which data is expressed as sets of curves instead  of discrete numerical values as in traditional statistics. Techniques of FDA have been successfully applied to a variety of problems in biomedicine and public health \cite{ullah2013applications}. In a recent paper, functional principal components analysis (fPCA) combined with functional clustering was used to identify patterns of COVID-19 incidence and mortality across countries \cite{Carroll2020,kumar2020express}. 
Further, variations of subset selection problems in functional contexts have also been addressed recently, such as regression analysis with a scalar response and a functional predictor \cite{james2009functional}, dimension reduction of a functional predictor for a categorical variable \cite{tian2013interpretable}, and others \cite{fraiman2016feature,cuevas2014partial}. 
Thus, the present study has the secondary goal of introducing the functional extension of the QR decomposition as an addition to the set of available FDA tools.
 
\section{Data}

\subsection{Description and pre-processing}

The evolution of the pandemic is assessed in terms of mortality rates (i.e., death counts per day), which provide a more reliable measure than infection rates \cite{vasconcelos2020modelling}. Official data of COVID-19 were obtained from a repository at the Ministry of Health of Brazil \cite{Health2020},  accessed on January 21, 2022. The data consists of records of deaths counts per day since February 25, 2020, in Brazil's 27 federative units (26 states and a Federal District). For simplicity, the federative units will be be called ``states'' throughout the analysis.  

For each state, the period from the first confirmed death was extracted, and all extracted records were cut to the length of the shortest one (646 days). Then, the records were normalized to population size of each state and expressed in deaths per million individuals,
\begin{equation}
x_{ij} = \frac{\text{Number of deaths at day }j}{\text{Population size}} \times 10^6
\end{equation}
for $i=1, 2,\ldots, 27$ and $j = 1, 2,\ldots, 646$.

A few isolated mortality values were detected in the records, and those were removed by averaging them with nearby data points, as follows: if $x_{ij} < 0$, then 
\begin{equation}
x_{ik} = \frac{1}{3}\sum_{\ell=-1}^1 x_{i,j+\ell}
\end{equation} 
for $k=j-1, j, j+1$.

In addition, a square root transformation $y_{ij}=\sqrt{x_{ij}}$ was applied to the data. The transformation compresses the dynamic range of the data, which prevents the occurrence of negative values of death rates when reconstructing the data from the selected subset \cite{Mokhtari2007}. A logarithmic transformation has the same effect and was also tested, but it tended to produce larger errors.

\subsection{Functional form}

The first step of the analysis is to put the discrete data into functional form \cite{Ramsay1997}. 

For each state $i$, the existence of a smooth non-negative real function $f_i(t)$ is assumed, such that  
\begin{equation}
y_{ij} = f_i(t_j)+\varepsilon_{ij},
\end{equation}
where $t_j$ is the time at the end of day $j$ (with $t_1=0$), and $\varepsilon_{ij}$ is an observational error or noise term.  
Each mortality function $f_i$ is defined over the domain $t\in [0, T]$, with $T= 345$ days, and is expressed in a basis expansion form
\begin{equation}
f_i(t)=\sum_{k=1}^Kc_{ik}g_k(t)
\label{basis}
\end{equation}
where $g_k(t)$, $k=1, 2,\ldots, K$ is a set of basis functions and $c_{ik}$ are the expansion coefficients. The expansion coefficients are computed by minimizing the cost function
\begin{equation}
F_{\lambda, f_i}=\sum_i \left\{ \sum_j\left[y_{ij}-f_i(t_j)\right]^2 +\lambda\int_T\left[D^2f_i(t)\right]^2dt\right\},
\label{cost}
\end{equation}
where $\lambda$ is a roughness penalty coefficient and $D^2$ denotes the second order derivative.

For the basis in Eq.~\ref{basis}, a truncated Fourier cosine series \cite{davis1989} was adopted, i.e.,
\begin{align}
g_1(t) & = 1/\sqrt{T},\\
g_k(t) & = \sqrt{2/T} \cos k\pi t/T, \quad k=2, 3,\ldots, K.
\end{align}
This basis was chosen because of its stability, ease of computation, and orthonormality on the interval $[0, T]$, which facilitates the QR decomposition. 
 A basis size of $K=20$ was selected by visual inspection of the results. Further, the optimal roughness penalty coefficient $\lambda$ was determined by minimizing the sum of the generalized cross validation measure (GCV) for each $f_k$ function \cite{Ramsay1997}, which produced $\lambda = 10$.

\begin{figure}
\centering
\includegraphics{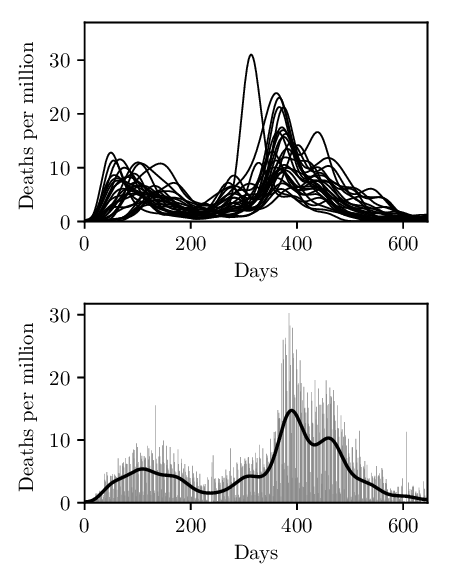}
\caption{Daily deaths per million inhabitants. Top: functional data. Bottom: original data for the state of Minas Gerais (gray bars) and functional form (black curve).}\label{fData}
\end{figure}

Fig.~\ref{fData} shows all data in functional form  and one example comparing the functional form to the original discrete data. The resultant functions are visually smooth and approximate well the original data, without weekly or short-term fluctuations.

\section{The QR decomposition}

\label{QR} 

In the so-called subset selection problem of linear algebra, a data matrix $A\in\mathbb{R}^{m\times n}$ and an observation vector $b\in\mathbb{R}^{m\times 1}$ are given,  with $m\ge n$, and a predictor vector $x$ is sought in the least squares sense; i.e., a minimizer of $\|Ax-b\|_2^2$ \cite{Golub1996}. However, instead of using the whole data matrix $A$ to predict $b$, only a subset of its columns is used so as to filter out any data redundancy.
This problem may be solved by the QR decomposition with column pivoting. The decomposition expresses $A$ in the form $AP=QR$, where $P\in\mathbb{R}^{n\times n}$ is a column permutation matrix, $Q\in\mathbb{R}^{m\times m}$ is an orthogonal matrix, and $R\in\mathbb{R}^{m\times n}$ is an upper triangular matrix with positive diagonal elements. A simplified variant is the ``thin'' version, in which $Q\in\mathbb{R}^{m\times n}$  and $R\in\mathbb{R}^{n\times n}$.
 
The first column of $AP$ is the column of $A$ that has the largest 2-norm, and the $k$th column of $AP$ ($k>1$) is the column of $A$ with the largest component in a direction orthogonal to the directions of the first $k-1$ columns. 
Thus, the algorithm reorders the columns of $A$ so as to make its first columns as well conditioned as possible. The first columns of $AP$ may be then adopted as the sought subset of least dependent columns. 
The diagonal elements of $R$ ($r_{ii})$, also called the ``$R$ values'', measure the size of the orthogonal components, and they appear in decreasing order for $i=1,\ldots,n$.

The decomposition may be extended to the functional case as follows. First, the data set of $n$ functions $f_i(t)$ is expressed as $A=[f_1, f_2, \ldots, f_n]$. From Eq.~(\ref{basis}), we have
\begin{equation}
A=GC,
\label{basisM}
\end{equation}
where $G = [g_1, g_2,\ldots, g_K]$ and $C$ is a $K\times n$ matrix of coefficients $c_{ik}$. Letting $AP=QR$ and expressing functions $g_i(t)$ in the same basis system as functions $f_i(t)$, we have $Q=GB$, where $B$ is a $K\times n$ orthogonal matrix of coefficients $b_{ik}$. Replacing into Eq.~(\ref{basisM}) and simplifying, we obtain
\begin{equation}
CP = BR
\label{qr2}
\end{equation}
which represents the standard (discrete) QR decomposition of matrix $C$, and may be computed using available algorithms of matrix algebra.

Once a suitable number $k$ of independent mortality functions has been chosen, the data set is approximated as the linear combination of the first $k$ functions, with $A\approx GC'$, and 
\begin{equation}
C' = B_{Kk}R_{kn}P^T,
\label{qr3}
\end{equation}
where $B_{Kk}$ and $R_{kn}$ are formed by the first $k$ columns of $B$ and the first $k$ lines of $R$, respectively. Finally, the regions of influence of each independent mortality function is determined by the size of the elements of $C'$; i.e., element $c'_{ij}$ measures the relative effect of function $i$ over state or country $j$.

\section{Results of the COVID-19 data analysis}

\begin{figure}
\centering
\includegraphics{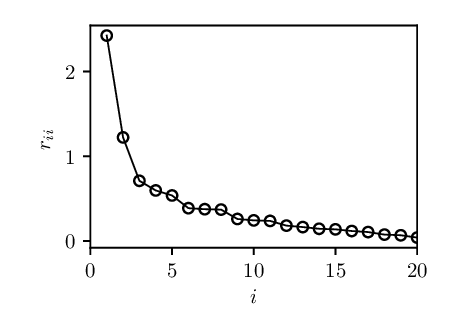}
\caption{$R$ values.}\label{rvalues}
\end{figure}

Fig.~\ref{rvalues} shows the whole set of $R$ values for the data. The $R$ values decrease as the number of selected functions increases, and their distribution suggest two main independent mortality functions \cite{setnes2001rule}.

\begin{figure}
\centering
\includegraphics{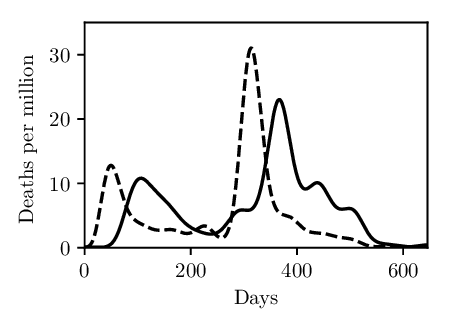}
\caption{Main independent mortality curves: Mato Grosso (solid curve) and  Amazonas (dashed curve).}\label{patterns}
\end{figure}

\begin{figure}
\centering
\includegraphics{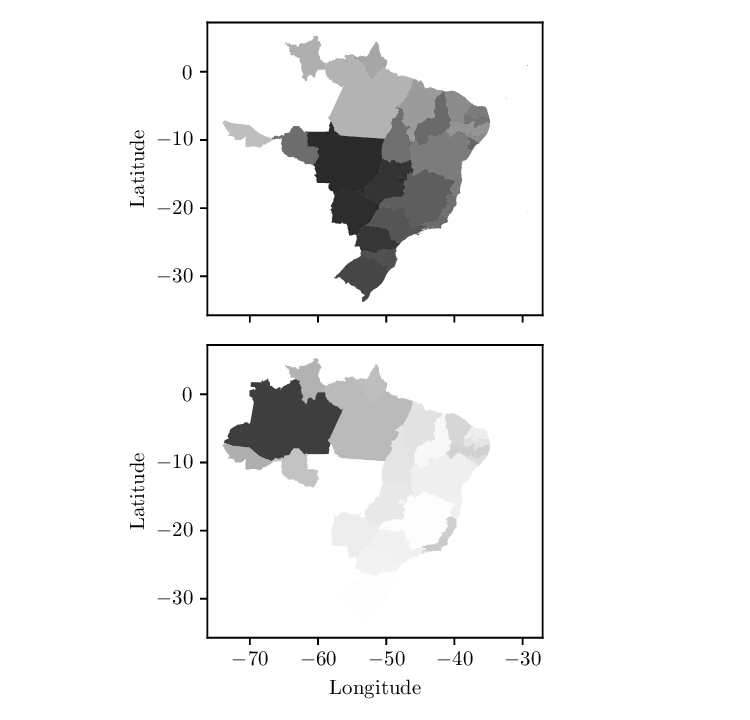}
\caption{Epidemiological regions defined by the mortality patterns of Mato Grosso (top) and Amazonas (bottom). The state corresponding to each pattern is the darkest one in each plot. The magnitude of the fit coefficient of each state is represented in gray scale.} \label{regions}
\end{figure}

The two main mortality functions correspond to the states of Mato Grosso (MT) in west-central Brazil, and Amazonas (AM) at the northwest, and they are plotted in Fig.~\ref{patterns}. Fig.~\ref{regions} shows the respective epidemiological regions that result from fitting the remaining states to the two main ones, as explained in Section \ref{QR}. The first region encompasses most of the country from the center and northeast to the south, whereas the second one contains the Amazonian region at the northwest.

Both curves in Fig.~\ref{patterns} have a similar two-peak pattern, with a 50-day shift between them. The first peaks correspond to the initial wave of the pandemic, and they occur in mid April 2020 in Amazonas and beginning of June in Mato Grosso. The earlier occurrence in Amazonas may be consequence of its international borders with Peru, Colombia and Venezuela, and the flow of people across them \cite{cimerman2020deep}. Other contributing factors may have been its high percentage of indigenous population which is more susceptible against contagious diseases, as well as its poorer developed public health care system \cite{palamim2020covid}.

The second peaks occurs at the beginning of 2021 in Amazonas and end of February in Mato Grosso. The timing matches the appearance of the new lineage P.1 of the SARS-CoV-2 virus, which had a higher transmissibility than previous lineages and was first detected in Manaus (capital city of Amazonas) \cite{sabino2021resurgence}.

\section{Conclusion}

This letter has introduced a simple functional extension of the QR decomposition technique of linear algebra, and shown its application to identify independent patterns of COVID-19 evolution in Brazil. Each pattern defines an epidemiological region, and the overall evolution of the pandemic in the country may be modeled (in the square root domain) as linear combination of the behavior of those regions. 
Naturally, the accuracy of the model depends on the number of independent patterns considered. Only the first two mortality patterns were discussed here for a general qualitative view; however, a larger number should be included if a more precise representation is desired.   

The functional expansion of the data adopted an orthogonal basis to facilitate the computation of the QR decomposition. Nevertheless, further development of the decomposition algorithm to allow for the use of non-orthogonal basis systems, such as the widely used B-splines, would be desired as a next step. 

\section*{Acknowledgments}

This work was supported by the Committee of Research, Innovation, and Extension to Combat COVID-19 (COPEI) of the University of Brasília.

\end{document}